# An experimental and theoretical investigation of the $N(^4S) + C_2(^1\Sigma_g^+)$ reaction at low temperature

*Jean-Christophe Loison,[a] Xixi Hu,[b] Shanyu Han,[b] Kevin M. Hickson,[a,]\* Hua Guo[c] and Daiqian Xie[b,d,]\**

**Corresponding Author**

\* Correspondence to: km.hickson@ism.u-bordeaux1.fr

Rate constants for the $N(^4S) + C_2(^1\Sigma_g^+)$ reaction have been measured in a continuous supersonic flow reactor over the range 57 K ≤ T ≤ 296 K by the relative rate technique employing the $N(^4S) + OH(X^2\Pi) \rightarrow H(^2S) + NO(X^2\Pi)$ reaction as a reference. Excess concentrations of atomic nitrogen were produced by the microwave discharge method and $C_2$ and OH radicals were created by the *in-situ* pulsed laser photolysis of precursor molecules $C_2Br_4$ and $H_2O_2$ respectively. In parallel, quantum dynamics calculations were performed based on an accurate global potential energy surfaces for the three lowest lying quartet states of the $C_2N$ molecule. The $1^4A''$ potential energy surface is barrierless, having two deep potential wells corresponding to the NCC and CNC intermediates. Both the experimental and theoretical work show that the rate constant decreases to low temperature, although the experimentally measured values fall more rapidly than the theoretical ones except at the lowest temperatures. Astrochemical simulations indicate that this reaction could be the dominant source of CN in dense interstellar clouds.

## 1.Introduction

Nitrogen containing compounds have been observed in a wide range of astronomical environments,

confirming that the chemistry of nitrogen is active throughout the Universe. In the dense interstellar medium, where temperatures are often lower than 10 K and photodissociation processes initiated by stellar photons play no part, ground state atomic N($^4$S) and molecular N$_2$($^1\Sigma_g^+$) are thought to represent the major reservoirs of elemental nitrogen. Nevertheless, nitrogen chemistry is not well understood, particularly at low temperature. Although N($^4$S) is an open-shell atomic radical, it generally displays low reactivity towards closed-shell molecules, in contrast to other abundant atomic radicals in the interstellar medium such as atomic carbon C($^3$P) and atomic oxygen O($^3$P). In addition, although ion-atom reactions with H$_3^+$ leading to hydride formation are important for C($^3$P) and O($^3$P), such reactions are prohibited for N($^4$S) due to an activation barrier for one pathway[1,2] leading to NH$_2^+$ + H products whilst another pathway leading to NH$^+$ + H$_2$ products is endothermic.[3] As dense clouds themselves evolve from diffuse ones which contain species in mostly atomic or ionic form, the chemistry of transformation from atomic to molecular nitrogen remains an unresolved issue. As N$_2$ is characterized by a strong covalent bond, it is difficult to incorporate nitrogen from N$_2$ into more complex molecules. Consequently, it is critically important to quantify the pathways leading to N$_2$ formation and the fraction of elemental nitrogen residing in this form to improve our overall understanding of the chemical evolution of interstellar media.

Two mechanisms, both involving neutral radical species are thought to govern N$_2$ formation; one through coupled nitrogen and oxygen chemistry (Mechanism A) and one through coupled nitrogen and carbon chemistry (Mechanism B):

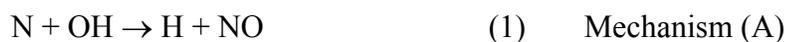
N + OH → H + NO         (1)     Mechanism (A)

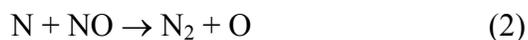
N + NO → N$_2$ + O         (2)

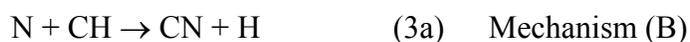
N + CH → CN + H         (3a)    Mechanism (B)

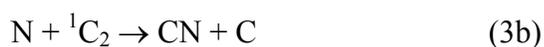
N + $^1$C$_2$ → CN + C         (3b)

N + C$_2$N → CN + CN     (3c)

N + CN → N$_2$ + C     (4)

Earlier theoretical and experimental studies of reactions (1), (2), (3a) and (4)[4-11] have shown that these individual processes are likely to be less efficient at low temperatures than previously thought. Moreover, as the major source of interstellar NO is reaction (1) and its major sinks are well known (NO is destroyed by reaction (2) in addition to the C + NO reaction) we can quantitatively assess N$_2$ production by mechanism (A). In contrast, Daranlot et al.[7] recently demonstrated that a tenfold decrease of the rate constant for reaction (3a) with respect to earlier estimations led to only a minor reduction in the overall CN abundance in a dense cloud model, despite the fact that this reaction was previously considered to be a major source of CN.[12] This finding indicates that other sources of CN radicals should be investigated to evaluate their potential for CN formation. One other reaction which might be a significant source of CN in interstellar clouds is reaction (3b). Although no previous kinetic studies exist, the vibrational distribution of nascent ground state CN(X$^2\Sigma^+$) produced by reaction (3b) has been investigated experimentally above room temperature.[13,14] These authors studied the decay of excited vibrational states of CN(X$^2\Sigma^+$) in their flow reactor as a function of time. The value of the rate constant for the N + C$_2$ reaction at room temperature (where C$_2$ refers to a mixture of both ground singlet state and first excited triplet state C$_2$, hereafter denoted $^1$C$_2$ and $^3$C$_2$ respectively) was estimated to be around 2 × 10$^{-10}$ cm$^3$ molecule$^{-1}$ s$^{-1}$,[13] indicating that reaction (3b) is likely to be barrierless and therefore potentially rapid below room temperature. Current astrochemical databases[15] use a rate constant value of 5.0 × 10$^{-11}$ cm$^3$ molecule$^{-1}$ s$^{-1}$ independent of temperature for the N + $^1$C$_2$ reaction. Whilst the rate of the N + $^1$C$_2$ reaction has never been studied either theoretically or experimentally, an experimental study of the N + $^3$C$_2$ reaction has been performed at room temperature by Becker et al.,[16] using an optical titration method to determine the excess atomic nitrogen concentration. A value of 2.8 × 10$^{-11}$ cm$^3$ molecule$^{-1}$ s$^{-1}$ was obtained for the rate constant, much slower than the estimated value for the corresponding $^1C_2$ reaction; a result which is in line with previous work on the relative reactivity of singlet versus triplet $C_2$.[17,18]

The present study reports the results of kinetic measurements of the reaction of $N(^4S)$ with $^1C_2$ to temperatures as low as 57 K. The experiments are complemented by quantum scattering calculations based on newly developed accurate potential energy surfaces (PESs) for the three lowest lying quartet states of the NCC/CNC intermediate complex, allowing temperature dependent rate constants to be calculated. The experimental and theoretical methodologies used to investigate this reaction are outlined in section 2. Section 3 presents a comparison of the experimental and theoretical results whilst the conclusions and astrophysical implications are developed in section 4.

## 2. Methodology

**Experimental methods**

Experiments were performed in a continuous supersonic flow reactor which has been described previously.[5,19] The reactor is based on the use of axisymmetric Laval nozzles whose convergent-divergent profiles are carefully designed to provide cold supersonic flows of uniform density and velocity at a given temperature for a specified carrier gas. Four different nozzles were used in the present study, allowing us to perform kinetic measurements at temperatures of 57 K, 81 K, 146 K and 167 K. In addition, experiments were performed at 296 K by removing the nozzle and by reducing the flow velocity, effectively allowing the reactor to be used as a slow flow flash photolysis system. Atomic nitrogen was produced upstream of the Laval nozzle through the microwave induced dissociation of $N_2$ carried by a secondary argon flow. For this purpose, a Vidal type cavity operating at 2.45 GHz and up to 200 W was mounted on the sidearm of a Y shaped quartz inlet tube leading to the nozzle reservoir. Atomic nitrogen was produced mostly in the ground $N(^4S)$ state and excited states $N(^2P^0)$ and $N(^2D^0)$ were either quenched or removed by reaction before reaching the cold supersonic flow, as shown by an earlier study under similar conditions[5] to the present experiments. As the high discharge power also resulted in elevated temperatures upstream of the Laval nozzle, a type K thermocouple was used to record the reservoir temperature in separate calibration experiments using identical conditions to the main experiments. The impact and stagnation pressures were also recorded during these calibration experiments allowing us to calculate the supersonic flow temperatures, densities and velocity profiles.

Rate constants for reaction (3b) were measured by the relative rate method, using the previously measured rates of the N + OH reaction as the reference.[5] OH radicals in the $X^2\Pi_i$ state were generated by the *in-situ* pulsed photolysis of $H_2O_2$ with a 10 Hz frequency quadrupled Nd:YAG laser at 266 nm with ~ 27 mJ of pulse energy. $H_2O_2$ was introduced into the reactor by bubbling a small flow of carrier gas through a 50% weight mixture of $H_2O_2$ / $H_2O$. An upper limit of $2.3 \times 10^{12}$ molecule cm$^{-3}$ was estimated for the gas phase concentration of $H_2O_2$ in the supersonic flows from its saturated vapour pressure, providing OH concentrations lower than $2.3 \times 10^{10}$ molecule cm$^{-3}$ from calculations of the $H_2O_2$ photodissociation efficiency. $C_2$ molecules in both the $^1\Sigma_g^+$ and $^3\Pi_u$ states (separated by only 610 cm$^{-1}$ considering zero point energy) were generated by the *in-situ* pulsed (multi)photon dissociation of $C_2Br_4$ using the same frequency quadrupled Nd:YAG laser. Solid $C_2Br_4$ which has an estimated saturated vapour pressure of 82 mTorr at 298 K was held at ambient temperature in a separate vessel. A small gas flow was used to carry the $C_2Br_4$ vapour into the reactor. An upper limit of $1.0 \times 10^{11}$ molecule cm$^{-3}$ was estimated for the gas phase concentration of $C_2Br_4$ in the chamber.

The probe laser system for the detection of OH radicals by laser-induced fluorescence (LIF) has been described previously[20] with radical fluorescence being monitored using a UV sensitive photomultiplier tube (PMT) coupled with a narrowband (10 nm FWHM) interference filter centered on 310 nm and a boxcar integration system. $^1C_2$ radicals were also followed by LIF through the ($D^1\Sigma_u^+$, v = 0 ← $X^1\Sigma_g^+$, v = 0) transition. Tunable radiation around 231 nm was produced in a similar manner to the study of Daugey et al.[18] Fluorescence emission from the $D^1\Sigma_u^+$ state with a lifetime of approximately 15 ns was observed on-resonance using a second UV sensitive PMT coupled with a fast preamplifier, a narrowband (10 nm FWHM) interference filter and a second boxcar integrator. The probe lasers were coaligned and couterpropagated with respect to the photolysis laser along the supersonic flow axis. $^3C_2$ radicals were not monitored in this study. Instead, test experiments, which are outlined in the results and discussion section, were performed to ensure that these radicals did not interfere with our investigation of the kinetics of the N + $^1C_2$ reaction.

All gases were flowed directly from cylinders with no further purification prior to usage (Linde: Ar 99.999%, $O_2$ 99.999%; Air Liquide: $N_2$ 99.999%). The carrier gas and precursor flows were all passed into the reservoir via digital mass flow controllers. The controllers were calibrated using the pressure rise at constant volume method for the specific gas used.

**Theoretical methods**

The reaction between N($^4S$) and $C_2$($X^1\Sigma_g^+$) leading to the products C($^3P$) and CN($X^2\Sigma^+$) is associated with the lowest quartet electronic state ($a^4\Sigma^-$ for $C_{\infty v}$, $1^4A''$ for $C_s$).[13] In this work, a new *ab initio* PES of the $1^4A''$ state has been developed at the Davidson corrected multi-reference configuration interaction (MRCI + Q) level[21-23] using the augmented correlation consistent polarized valence triple zeta (aug-cc-pVTZ) basis set. The state-averaged completed active space self-consistent field (SA-CASSCF) calculations including three states ($1^4A'$, $1^4A''$ and $2^4A''$) were performed to provide a balanced description of the $1^4A''$ PES, because the $1^4A'$ and $2^4A''$ states are degenerate at the NCC linear structure. The CASSCF calculations were performed with a full-valence active space comprising 13 electrons in 12 orbitals [i.e., CAS (13,12)], while the three 1s orbitals were doubly occupied and fully optimized. In the subsequent MRCI calculations, only 11 valence electrons in 11 orbitals were included in the active space, while all three 1s atomic orbitals and one 2s orbital of the nitrogen atom were frozen. All *ab initio* calculations were carried out with the MOLPRO 2010 package.[24]

The PES was spline fitted based on about 13,000 *ab initio* points which were calculated on a grid defined by $1.7 \leq R_{NC}/a_0 \leq 15.0$, $1.75 \leq R_{CC}/a_0 \leq 14.0$ and $0 \leq \theta_{N-C-C}/\text{deg} \leq 180$. All stationary points on the PES between reactants and products are listed in Table 1, with comparison with previous theoretical results.[25-27] The reaction path is barrierless and has two deep wells corresponding to the NCC and CNC species, connected by two transition states and a cyclic intermediate, as shown in Fig. 1(a). It can be seen from Fig. 1(c) that there are a shallow van der Waals well and a submerged "reef" structure below the dissociation limit at the N + $C_2$ asymptotic region. The product asymptote is 1.374 eV lower than the reactant asymptote without considering zero point energies. The experimental exoergicity (1.48 ± 0.07) eV was estimated by the difference between the dissociation energies of $C_2(X^1\Sigma_g^+)$ and $CN(X^2\Sigma^+)$, which are (6.29 ± 0.02)[28] and (7.77 ± 0.05) eV,[29] respectively.

The reactive scattering dynamics of the title reaction was investigated using the Chebyshev real wave packet (CRWP) method.[30] Here, the reactant Jacobi coordinates $(R, r, \gamma)$ are used, which represent the distance between the N atom and the $C_2$ centre, the $C_2$ bond length and the angle between $R$ and $r$, respectively. The Hamiltonian was given by ($\hbar = 1$):

$$\hat{H} = -\frac{1}{2\mu_R}\frac{\partial^2}{\partial R^2} - \frac{1}{2\mu_r}\frac{\partial^2}{\partial r^2} + \frac{(\hat{J}-\hat{j})^2}{2\mu_R R^2} + \frac{\hat{j}^2}{2\mu_r r^2} + V(R,r,\gamma) \quad (5)$$

where $\mu_r$ and $\mu_R$ are the reduced masses of the diatomic $C_2$ and N-$C_2$, respectively. $j$ and $J$ stand for the diatomic $C_2$ and total angular momenta.

The Hamiltonian and wave packet were discretized in a mixed basis/grid representation.[31] To make the calculations efficient, an $L$-shaped grid was used for the radial coordinate.[32] For each given total angular momentum $J$ and parity $p$, the wave function in the body-fixed (BF) frame can be expanded as:

$$\left|\Psi^{Jp}\right\rangle = \sum_{\alpha_r \alpha_R jK} C_{\alpha_r \alpha_R jK}^{Jp} \left|\alpha_r\right\rangle \left|\alpha_R\right\rangle \left|jK,Jp\right\rangle \quad (6)$$

where $\alpha_r$ and $\alpha_R$ denote the grid indices for the two radial coordinates and $K$ is the projection of $J$ onto the BF $z$ axis. The parity-adapted angular basis can be written as

$$\left|jK;Jp\right\rangle = (2+2\delta_{K,0})^{-1/2}(\left|JK\right\rangle\left|jK\right\rangle + p(-1)^J \left|J-K\right\rangle\left|j-K\right\rangle) \quad (7)$$

where $\left|jK\right\rangle \equiv Y_j^K(\gamma,0)$ are normalized associated Legendre functions and $\left|JK\right\rangle \equiv \sqrt{(2J+1)/8\pi^2} D_{KM}^J$ are

normalized Wigner ($D_{KM}^{J}$) rotation matrices.[33]

The initial wave packet $|\Psi_i\rangle$ in the space-fixed frame was defined below:[30]

$$|\Psi_i\rangle = Ne^{-(R-R_0)^2/2\delta^2} \cos k_0 R |\varphi_{v_i j_i}\rangle |j_i l_i, Jp\rangle \quad (8)$$

Where $N$ is the normalization factor, $R_0$ and $\delta$ represent respectively the initial position and width of the Gaussian function, $k_0 = \sqrt{2\mu_R E_0}$ is the mean momentum and $|\varphi_{v_i j_i}\rangle$ ($v_i=0$, $j_i=0$) is the eigenfunction of the initial rovibrational state of $C_2$. The initial wave function was transformed into the body-fixed frame before propagation.

We used a coordinate transformation method[32] to carry out final state projection in the product channel, which yields the $S$-matrix elements ($S_{v_f j_f K_f \leftarrow v_i j_i K_i}^{J}(E)$) and the state-to-state integral cross sections (ICSs):

$$\sigma_{v_f j_f \leftarrow v_i j_i}(E) = \frac{\pi}{(2j_i+1)k_{v_i j_i}^2} \sum_{K_i} \sum_{K_f} \sum_{J} (2J+1) \left| S_{v_f j_f K_f \leftarrow v_i j_i K_i}^{J}(E) \right|^2 \quad (9)$$

in which $k_{v_i j_i}$ is the initial translational wave vector.

Finally, the initial state specified rate constant was given by a Boltzmann average of ICSs over the collision energy:

$$k_{v_i j_i}(T) = \frac{f_e(T)}{k_B T}\left(\frac{8}{\pi\mu_R k_B T}\right)^{1/2} \sum_{v_f j_f} \int_0^\infty \sigma_{v_f j_f \leftarrow v_i j_i}(E_c) e^{-E_c/k_B T} E_c dE_c \quad (10)$$

where $T$ is the temperature, $k_B$ is the Boltzmann constant, and the electronic statistical weighting factor $f_e(T)$ is 1.

The optimal numerical parameters listed in Table 2 were chosen by extensive convergence tests. Partial waves up to $J = 94$ were included in our calculation to converge the ICSs to 0.16 eV and the rate constants to 400 K. For $J \leq 70$, a maximal $K$ value of 30 converged the probabilities well, and $K_{max}=45$ was sufficient for $70 < J \leq 94$.

## 3. Results and Discussion

OH and $^1C_2$ radical fluorescence signals were recorded simultaneously as a function of time for all experiments. Atomic nitrogen was held in large excess with respect to the OH and $C_2$ radical concentrations so that simple exponential fits to the OH and $^1C_2$ temporal profiles yielded the pseudo-first order rate constants for the N + OH and N + $^1C_2$ reactions, $k'_{N+OH}$ and $k'_{N+C2}$ respectively. Examples of such traces are presented in Figure 2.

It can be seen that $^1C_2$ radicals react much more rapidly than OH radicals with atomic nitrogen at both temperatures, although the difference is clearly greater at higher temperature indicating that the rate constants for these two processes converge as the temperature falls. This measurement was repeated at several different atomic nitrogen concentrations with each nozzle by varying the $N_2$ flow through the microwave discharge and the microwave power. The values of $k'_{N+C2}$ obtained in this way for a range of atomic nitrogen concentrations were then plotted as a function of the corresponding values of $k'_{N+OH}$ as shown in Figure 3 yielding straight lines for all experiments and at all temperatures. Weighted linear least squares fits to the data yielded the ratio of the two rate constants $k'_{N+C2}$ / $k'_{N+OH}$ at a given temperature from the slopes.

One potential source of error in these measurements is the presence of $^3C_2$ in the supersonic flow. As the singlet and triplet states of $C_2$ are close in energy, there is a significant possibility that intersystem crossing could occur either through collisions with the carrier gas or with the atomic nitrogen coreagent. Reisler et al.[34] performed a series of experiments to determine the intersystem crossing rate constants for $^3C_2$ to $^1C_2$ transfer at 296 K with several colliders. They determined an upper limit for transfer by $N_2$ and Ar of $3 \times 10^{-14}$ $cm^3$ $molecule^{-1}$ $s^{-1}$ which translates to a maximum pseudo-first-order transfer rate of approximately 4900 $s^{-1}$ in our experiments at 296 K. As the buffer gas flow does not change for any single series of experiments, population transfer to the ground singlet state would result in a change in the y intercept value of the resulting second-order plot without changing the slope.

Nevertheless, as the atomic nitrogen concentration is varied by changing the $N_2$ flow into the microwave discharge, the measured $k'_{N+C2}$ values could still be affected by relaxation with $N_2$. Assuming a constant value for the $^3C_2$ to $^1C_2$ relaxation rate constant as a function of temperature, we estimate a maximum pseudo-first-order transfer rate of approximately 350 s$^{-1}$, a negligible quantity with respect to the overall $^1C_2$ decay rates.

Becker et al.[16] studied the loss of $^3C_2$ radicals in the presence of atomic nitrogen, measuring a rate constant for this process of $(2.8 \pm 1.0) \times 10^{-11}$ cm$^3$ molecule$^{-1}$ s$^{-1}$. As they did not attempt to detect the reaction products, it is possible that some or all of the observed decay of $^3C_2$ could have been due to population transfer to the ground singlet state. To test the idea that a non-reactive quenching mechanism for collisions with atomic nitrogen itself might influence the kinetic decay profiles, we performed a series of experiments whereby $^3C_2$ was rapidly removed from the cold supersonic flow by adding an excess of $O_2$. Paramo et al.[35] measured rate constants at low temperature for the reactions of both $^3C_2$ and $^1C_2$ with $O_2$. They showed that intersystem crossing and / or reaction with $^3C_2$ occurred over the range 24 – 300 K and that $^1C_2$ could be efficiently removed above 150 K using a large excess of $O_2$. Moreover, Le Picard et al.[36] have shown that the third-body reaction of OH with $O_2$ at 57 K is slow (leading to expected first-order decays due to $HO_3$ formation of around 200 s$^{-1}$ under these experimental conditions) so that this process should not interfere with the relative rate measurement of the N + $^1C_2$ reaction. In the present experiments, $O_2$ concentrations of $6.1 \times 10^{15}$ and $6.5 \times 10^{15}$ molecule cm$^{-3}$ were added to the flow at 57 K and 296 K respectively leading to estimated pseudo-first-order rate constants for $^3C_2$ loss of approximately $6 \times 10^4$ s$^{-1}$ and $2 \times 10^4$ s$^{-1}$. $^1C_2$ radicals were followed by LIF in the usual manner. The 57 K experiments performed in the presence and absence of $O_2$ yielded relative rate constants of $2.80 \pm 0.26$ and $2.28 \pm 0.20$ respectively. Although these values are slightly outside the combined error bars of the two experiments, the difference almost certainly reflects the measurement accuracy at this temperature, rather than a problem associated with quenching of $^3C_2$, given the large value for the pseudo-first-order decay rate with $O_2$.

At 296 K, the situation is somewhat different as it was not possible to add as much $O_2$ to the flow given the relatively high reactivity of $^1C_2$ with $O_2$ at this temperature. As a result, $^3C_2$ could potentially decay to the ground state on the same timescale as the main experiments. The second-order plots obtained for experiments performed at 300 K in the presence and absence of $O_2$ are shown in Figure 4. Whilst the $^1C_2$ loss rates are substantially larger in the presence of $O_2$ (given the fast supplementary loss for $^1C_2$ with $O_2$ of approximately $2 \times 10^4$ s$^{-1}$), the slopes are seen to be almost identical, indicating that intersystem crossing from $^3C_2$ to $^1C_2$ through collisions with atomic nitrogen plays little or no role here.

Temperature dependent rate constants for the N + $^1C_2$ reaction were obtained by multiplying the ratios obtained from the slopes of plots similar to the ones presented in Figures 3 and 4 by the value of the rate constant previously obtained for reaction (1) ($k_1$(56-300 K) = (4.5 ± 0.9) × 10$^{-11}$ cm$^3$ molecule$^{-1}$ s$^{-1}$).[5] These values are listed in Table 3 alongside other relevant information and are displayed as a function of temperature in Figure 5 alongside the present theoretical results.

A simple A × (T/300)$^B$ fit to the present experimental data yields a temperature dependence B = (0.58 ± 0.02) with A = (3.01 ± 0.30) × 10$^{-10}$ cm$^3$ molecule$^{-1}$ s$^{-1}$. Using this formula to extrapolate the data to lower temperature allows us to estimate a value for $k_{3b}$(10 K) = 4.2 × 10$^{-11}$ cm$^3$ molecule$^{-1}$ s$^{-1}$ which agrees well with the temperature independent value of 5.0 × 10$^{-11}$ cm$^3$ molecule$^{-1}$ s$^{-1}$ recommended by current astrochemical databases.[12-14] Whilst the agreement is good at 10 K, the values clearly diverge at higher temperature, leading to a sixfold underestimation of the rate constant at room temperature. While the computed and experimental rate constants are in excellent agreement at room temperature, the calculated rate is somewhat higher than the experimental rate at intermediate temperatures.

The discrepancy between the experimental and computed rate constants might originate from several possible sources. Firstly, the error could stem from the experimental measurements. As the rate constant is measured relative to that of the N + OH reaction (which was itself measured relative to the N + NO reaction),[5] there is a possibility that errors in these earlier investigations could propagate into the present study. A more detailed discussion of such problems can be found in Daranlot et al.[7] Secondly, the error could stem from the *ab initio* calculations. Despite the state of the art nature of the MRCI calculations reported here, a relatively small active space was used, because of the prohibitive computational costs for this extremely demanding system. This could introduce systematic errors, particularly in the entrance channel region. Secondly, due to difficulties in damping the wave packet with long de Broglie wavelengths, reaction probabilities may not converge completely at very low collision energies ($E_c \leq 0.02$ eV), which might affect the rate constant at very low temperatures. Finally, the non-adiabatic couplings between the $1^4A''$ and $2^4A''$ states which have not been considered in this work could potentially interfere.

## 4. Conclusions and Astrophysical Implications

Rate constants for the $N(^4S) + C_2(X^1\Sigma_g^+)$ reaction by both experimental and theoretical methods are reported here for a wide range of temperatures, extending down to 10 K. The reaction is shown to proceed through a complex forming mechanism involving the lowest lying quartet state ($a^4\Sigma^-$ for $C_{\infty v}$ symmetry, $^4A''$ for $C_s$ symmetry) of CCN/CNC leading to the formation of ground state products $C(^3P)$ and $CN(X^2\Sigma^+)$. The experimental and theoretical rate constants both increase as a function of temperature leading to a positive temperature dependence for the rate. Nevertheless, the experimental rate constants fall more rapidly than the calculated ones at low temperature so that the temperature dependence of the experimental rate is only partially reproduced by theory. Indeed, the $T^{0.6}$ dependence for the experimental rate of the $N + {}^1C_2$ reaction is the strongest yet obtained for any of the atomic nitrogen – molecular radical reactions to have been investigated experimentally.

In a recent paper, Daranlot et al.[7] demonstrated that a tenfold decrease of the rate constant for reaction (3a) with respect to earlier estimations led to only a minor reduction in the overall dense cloud abundance of CN radicals in a gas-grain astrochemical model, suggesting that other reactions could also contribute to CN formation. As current estimates of the rate constant for the N + $C_2$ reaction at 10 K are in reasonably good agreement with the value determined here, we can look in more detail at the results of these earlier simulations to estimate the relative importance of individual CN formation mechanisms. A more detailed description of the model parameters can be found in Daranlot et al.[7] Using the reactive flux as a guide (where the reactive flux = $k_{3b}$ [A][B], with $k_{3b}$ the rate constant at 10 K, and [A] and [B] the concentrations of the two reactants), for models employing two values of the initial carbon to oxygen abundance ratio (C/O = 0.7 and 1.2 with C = $1.7 \times 10^{-4}$ with respect to total hydrogen nH + 2nH$_2$) we observe that reaction (3b) is indeed the major source of CN radicals at early and intermediate times (< $1 \times 10^5$ years). For a C/O ratio of 1.2, reaction (3b) remains the dominant neutral CN formation mechanism over the entire simulated lifetime of the dense cloud although the HCNH$^+$ electronic recombination reaction is the most important source of CN radicals after $1 \times 10^5$ years. For the model employing a C/O ratio of 0.7, reaction (3a) competes with reaction (3b) after $1 \times 10^5$ years as the main neutral source of CN.

It should be noted that the relevance of certain CN production mechanisms can be underestimated using the reactive flux as the only criterion for classifying reactions. Indeed, the reaction between N and C$_2$N (reaction 3c) with an estimated rate constant at 10 K[37] of $5.0 \times 10^{-11}$ cm$^3$ molecule$^{-1}$ s$^{-1}$ is expected to be twice as important as its flux would suggest as it leads to the formation of two CN radicals per reactive event. Future experimental work will focus on the investigation of the kinetics of this reaction at low temperature to evaluate its contribution to interstellar CN production.

**Acknowledgements**

JCL and KMH are supported by the INSU-CNRS national programs PCMI and PNP and the Observatoire Aquitain des Sciences de l'Univers. XH and DX are supported by the National Natural

Science Foundation of China (21133006, 21273104, and 91021010) and the Ministry of Science and Technology (2013CB834601). HG acknowledges support from the DOE (DE-FG02-05ER15694).

[a]*Université de Bordeaux, Institut des Sciences Moléculaires, CNRS UMR 5255, F-33400 Talence, France. Tel: +33 5 40 00 63 42. E-mail: km.hickson@ism.u-bordeaux1.fr*

[b]*Institute of Theoretical and Computational Chemistry, Key Laboratory of Mesoscopic Chemistry, School of Chemistry and Chemical Engineering, Nanjing University, Nanjing 210093, China. E-mail: dqxie@nju.edu.cn*

[c]*Department of Chemistry and Chemical Biology, University of New Mexico, Albuquerque, NM 87131, USA. E-mail: hguo@unm.edu*

[d]*Synergetic Innovation Center of Quantum Information and Quantum Physics, University of Science and Technology of China, Hefei, Anhui 230026, China*

**Table 1** Comparison of geometries (in bohr and degree) and relative energies (in eV) of the stationary points with previous theoretical results.

| Stationary points | Method | $R_{NC}$ | $R_{CC}$ | $\theta_{N-C-C}$ | $\Delta E$ |
|---|---|---|---|---|---|
| N + CC | MRCI+Q/AVTZ | 15.0 | 2.36 | 0.0 | 0.0 |
| NCC minimum | MRCI+Q/AVTZ | 2.253 | 2.520 | 180.0 | -5.299 |
| | CASSCF/pVDZ[25] | 2.266 | 2.590 | 180.0 | -4.5 |
| | CASSCF/6s4p2d2f[26] | 2.246 | 2.574 | 180.0 | |
| | CCSD(T)/VQZ-F12[27] | 2.250 | 2.510 | 180.0 | |
| TS1 | MRCI+Q/AVTZ | 2.292 | 2.744 | 94.0 | -2.721 |
| INT | MRCI+Q/AVTZ | 2.419 | 2.615 | 73.0 | -3.138 |
| TS2 | MRCI+Q/AVTZ | 2.298 | 3.980 | 40.0 | -2.585 |
| CNC minimum | MRCI+Q/AVTZ | 2.354 | 4.708 | 0.0 | -3.992 |
| | CASSCF/pVDZ[25] | 2.394 | 4.788 | 0.0 | |
| CN + C | MRCI+Q/AVTZ | 2.221 | 20.0 | 0.0 | -1.374 |
| | Expt. | | | | -1.48 ± 0.07 |

**Table 2** Numerical parameters (atomic units) used in the wave packet calculations.

| | |
|---|---|
| Grid / basis ranges and sizes | $R \in [0.01, 19.0]$, $N_R^1 = 511$, $N_R^2 = 269$ |
| | $r \in [1.7, 15.7]$, $N_r^1 = 323$, $N_r^2 = 59$ |
| | $\gamma \in [0, 180°]$, $j = 0 \sim j_{max} = 300$, $N_j = 301$ |
| Projection | $R'_\infty = 12.0$ |
| Initial wave packet | $R_0 = 11.0$, $\delta = 0.1$, $E_0 = 0.05$ eV |
| Damping function | $D = \begin{cases} \exp[-0.01(\frac{R-14.0}{5.0})^2], & 14.0 \leq R \leq 19.0 \\ \exp[-0.09(\frac{r-14.9}{0.8})^2 - 0.007], & 14.9 < r \leq 15.7 \\ \exp[-0.007(\frac{r-13.9}{1.0})^2], & 13.9 \leq r \leq 14.9 \\ 1, & \text{otherwise} \end{cases}$ |
| Spectral range | Potential energy cutoff: 7.5 eV<br>Rotational kinetic energy cutoff: 5.5 eV |
| Propagation steps | 60000 |

**Table 3** Measured rate constants for the N + $^1C_2$ reaction

| $T$ / K | $[N]_{max}$ / $10^{14}$ atom cm$^{-3}$* | Flow density / $10^{16}$ cm$^{-3}$ | $N^\dagger$ | $k_{3b}$ / $10^{-11}$ cm$^3$ molecule$^{-1}$ s$^{-1}$ | $k'_{3b}/k'_1$ $\equiv k_{3b}/k_1$ |
|---|---|---|---|---|---|
| 57 ± 1 | 1.5 | 22.1 ± 0.6 (Ar) | 60 | 11.4 ± 2.7 | 2.54 ± 0.33 |
| 81 ± 2 | 2.2 | 13.0 ± 0.4 (Ar) | 58 | 14.5 ± 3.0 | 3.22 ± 0.14 |
| 146 ± 1 | 3.6 | 9.9 ± 0.4 (Ar) | 56 | 19.8 ± 4.1 | 4.40 ± 0.22 |
| 167 ± 1 | 2.0 | 8.5 ± 0.1 (N$_2$) | 60 | 20.9 ± 4.3 | 4.64 ± 0.16 |
| 296 | 2.0 | 16.2 (Ar) | 79 | 30.4 ± 6.4 | 6.76 ± 0.41 |

* Estimated from fits to the OH decay profiles.

$^\dagger$ Number of individual measurements

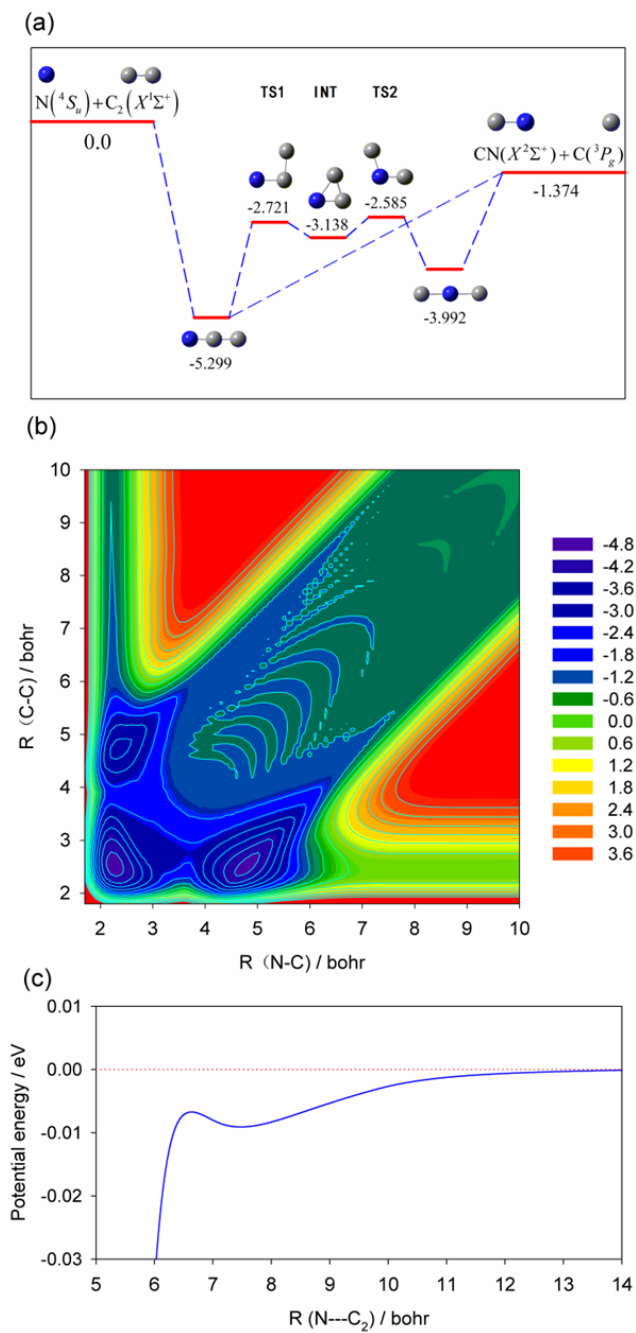

Fig. 1 (a) Potential energy diagram for the reaction N($^4$S) + C$_2$(X$^1\Sigma_g^+$) → C($^3$P) + CN(X$^2\Sigma^+$). The calculated energies (in eV) of the stationary points of NCC/CNC for the $1^4A''$ state are given in Table 1; (b) Contour plot of the NCC/CNC PES in internal coordinates. The energy zero is defined at the N + C$_2$ asymptote. (c) Minimal energy path in the N + C$_2$ asymptotic region.

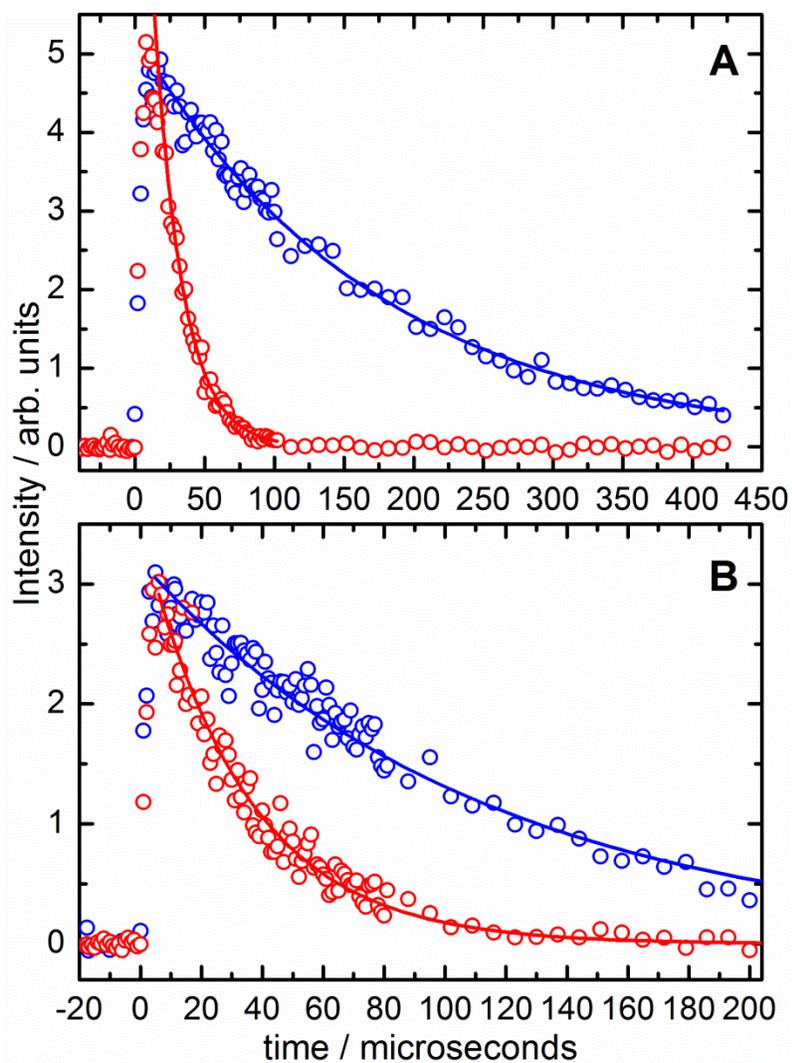

Fig. 2 Variation of the fluorescence emission signal from reactant $^1C_2$ radicals (red open circles) and OH radicals (blue open circles) as a function of time recorded simultaneously (A) at 296 K in the presence of an estimated excess of atomic nitrogen of $6.0 \times 10^{13}$ atom cm$^{-3}$. (B) at 81 K in the presence of an estimated excess of atomic nitrogen of $1.3 \times 10^{14}$ atom cm$^{-3}$. Only data points taken at times less than 100 μs were used to obtain the pseudo-first-order rate constant for $^1C_2$ in panel (A).

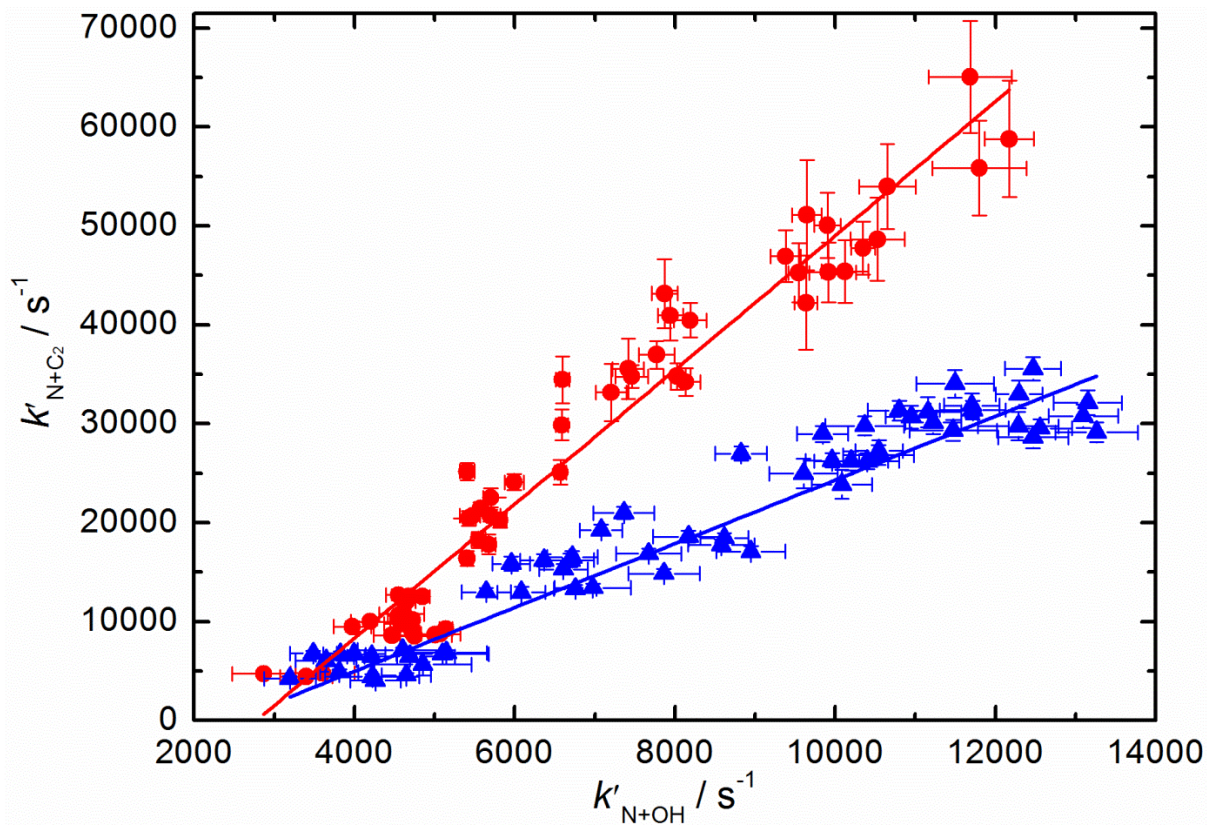

Fig. 3 Pseudo-first-order rate constants for reaction (3b) as a function of the pseudo-first-order rate constants for reaction (1) at 81 K (blue filled triangles) and 296 K (red filled circles). Weighted linear least squares fits yield the ratios of the second-order rate constants $k_{3b}/k_1$. The error bars on the ordinate reflect the statistical uncertainties at the level of a single standard deviation obtained by fitting to $^1C_2$ LIF profiles such as those shown in Fig. 2. The error bars on the abscissa were obtained in the same manner by fitting to the OH LIF profiles.

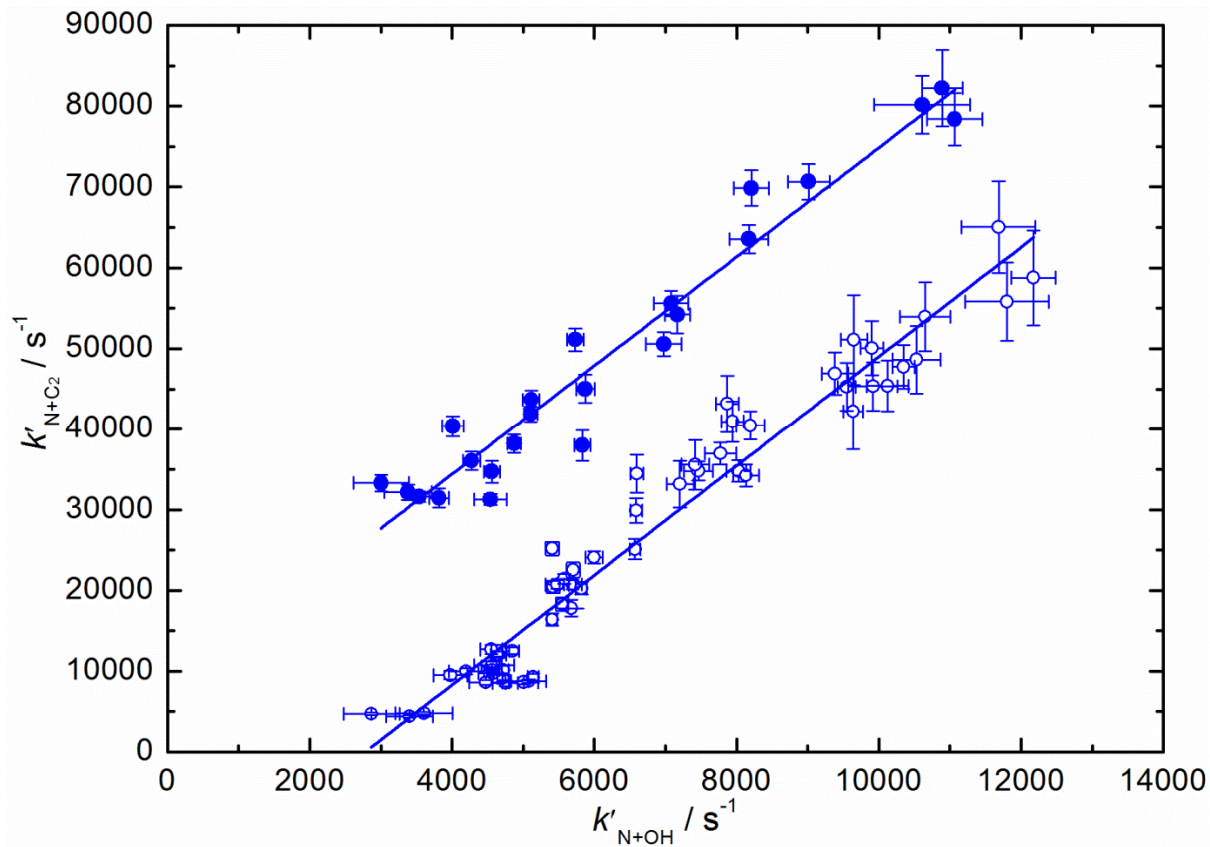

Fig. 4 Pseudo-first-order rate constants for reaction (3b) as a function of the pseudo-first-order rate constants for reaction (1) at 296 K. (Blue filled circles) with $[O_2] = 6.5 \times 10^{15}$ molecule cm$^{-3}$; (blue open circles) in the absence of $O_2$. Weighted linear least squares fits yield the ratios of the second-order rate constants $k_{3b}/k_1$. The error bars on the ordinate reflect the statistical uncertainties at the level of a single standard deviation obtained by fitting to $^1C_2$ LIF profiles such as those shown in Fig. 2. The error bars on the abscissa were obtained in the same manner by fitting to the OH LIF profiles.

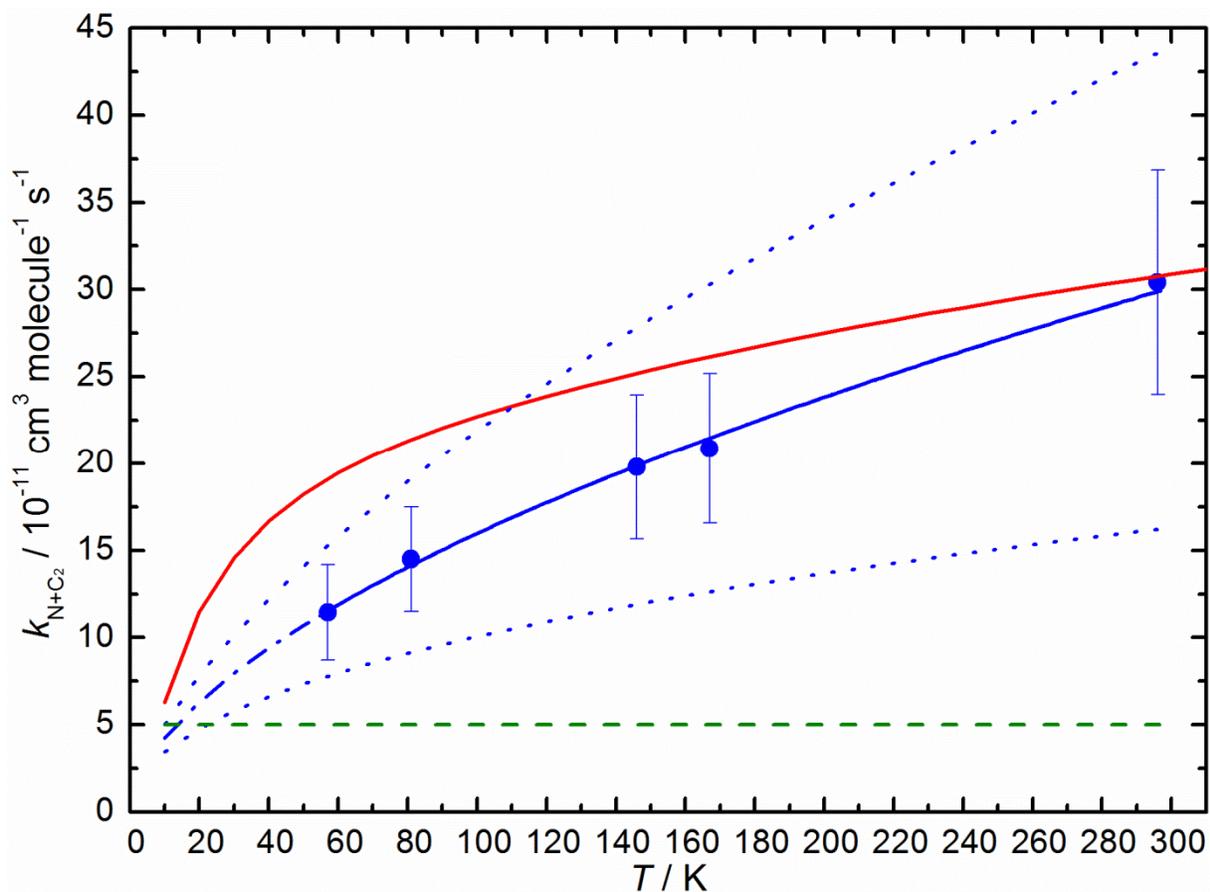

Fig. 5 Rate constants for the N($^4$S) + $^1$C$_2$ reaction as a function of temperature. (Blue filled circles) this experimental work using reaction (1) as a reference; (solid blue line + dashed dotted line) fit to the current experimental data + extrapolation to 10 K; (blue dotted lines) 95 % confidence limits from fitting to the present experimental data; (red solid line) this work, calculated rate constant for reaction (3b) over the $1^4A''$ surface; (green dashed line) currently recommended value from Wakelam et al.[15]